\documentclass[pra,superscriptaddress,showpacs,eqsecnum,amsmath,twocolumn]{revtex4}
\usepackage{graphicx}
\usepackage{bm}

\begin{document}
\newcommand{\ket}[1]{|{#1}\rangle}
\newcommand{\bra}[1]{\langle{#1}|}
\newcommand{\bracket}[2]{\langle{#1}|{#2}\rangle}
\newcommand{\tr}{\mathop{\mathrm{tr}}\nolimits}
\renewcommand{\Re}{\mathop{\mathrm{Re}}}
\renewcommand{\Im}{\mathop{\mathrm{Im}}}

\newcommand{\beq}{\begin{equation}}
\newcommand{\eeq}{\end{equation}}
\newcommand{\barr}{\begin{eqnarray}}
\newcommand{\earr}{\end{eqnarray}}
\newcommand{\andy}[1]{ }

\title{Optimization of a Neutron-Spin Test of the Quantum Zeno Effect}
\author{Paolo Facchi}
\email{paolo.facchi@ba.infn.it}
\affiliation{Dipartimento di Fisica, Universit\`a di Bari and
Istituto Nazionale di Fisica Nucleare, Sezione di Bari, I-70126 Bari, Italy}
\author{Yoichi Nakaguro}
\author{Hiromichi Nakazato}
\email{hiromici@waseda.jp}
\affiliation{Department of Physics, Waseda University, Tokyo 169-8555, Japan}
\author{Saverio Pascazio}
\email{saverio.pascazio@ba.infn.it}
\affiliation{Dipartimento di Fisica, Universit\`a di Bari
and Istituto Nazionale di Fisica Nucleare, Sezione di Bari, I-70126 Bari, Italy}
\author{Makoto Unoki}
\author{Kazuya Yuasa}
\email{yuasa@hep.phys.waseda.ac.jp}
\affiliation{Department of Physics, Waseda University, Tokyo 169-8555, Japan}

\begin{abstract}
A neutron-spin experimental test of the quantum Zeno effect (QZE)
is discussed from a practical point of view, when the nonideal
efficiency of the magnetic mirrors, used for filtering the spin
state, is taken into account. In the idealized case the number $N$
of (ideal) mirrors can be indefinitely increased, yielding an
increasingly better QZE\@. By contrast, in a practical situation
with imperfect mirrors, there is an optimal number of mirrors,
$N_\text{opt}$, at which the QZE becomes maximum: more frequent
measurements would deteriorate the performance. However, a
quantitative analysis shows that a good experimental test of the
QZE is still feasible. These conclusions are of general validity:
in a realistic experiment, the presence of losses and
imperfections leads to an optimal frequency $N_\text{opt}$, which
is in general finite. One should not increase $N$ beyond
$N_\text{opt}$. A convenient formula for $N_\text{opt}$, valid in
a broad framework, is derived as a function of the parameters
characterizing the experimental setup.
\end{abstract}
\maketitle

\section{Introduction}
If very frequent measurements are made on a quantum system in
order to ascertain whether it is still in the initial state, its
evolution is slowed down and eventually totally hindered in the
limit of infinite frequency. This is the quantum Zeno effect
(QZE) \cite{ref:QZE,ref:reviewQZE,ref:QZEMisraSudarshan}, that was
considered little more than a curiosity until the experimental
confirmations by Itano \textit{et~al.}\ \cite{Itano} (that
followed a theoretical proposal by Cook \cite{Cook}) and by
Raizen's group in Texas \cite{ref:NonExponentialExp}. This last
experiment has proved the existence of the QZE for \textit{bona
fide} unstable systems and the occurrence of the inverse QZE,
i.e., acceleration of decay by repeated (not extremely frequent)
measurements \cite{ref:IQZE}. The temporal behavior of quantum
mechanical systems and in particular the nonexponential features
at short times, on which QZE and inverse QZE hinge, are reviewed
in Ref.\ \cite{ref:reviewQZE}.

We are now going through a phase of experimental verification of
the QZE\@. It is therefore important to understand the physical
meaning of ``infinitely'' frequent measurements, focusing on
practical applications, imperfections of the apparatus and
experimental losses as well as theoretical bounds. Some of these
problems were tackled in Ref.\ \cite{ref:OnQZE}. In this article,
we reconsider a proposal of an experimental test of the QZE that
makes use of neutron spin \cite{ref:NeutronSpinTestQZE}. In view
of the recent progress in perfect crystal neutron-storage
technology \cite{ref:NeutronStorage1991,ref:RauchQZE}, it is
necessary to investigate the physical properties of a Zeno setup,
focusing in particular on practical limits.

In this article we will study the practical \textit{imperfections
in the spectral decomposition}. In a few words, a ``spectral
decomposition'' {\em \`a la} Wigner \cite{Wigner} is a unitary
process that associates additional degrees of freedom to different
values of the observable to be measured. In this sense, it yields
no wave-function collapse. It is known, and will be reviewed in
Sec.\ \ref{sec:QZEIdeal}, that a frequent series of spectral
decompositions is sufficient in order to obtain a
QZE \cite{ref:reviewQZE,ref:NeutronSpinTestQZE,%
ref:PetroskyTasakiPrigogine}.

In the proposed neutron-spin experimental test of the QZE
\cite{ref:NeutronSpinTestQZE}, the spectral decomposition is
realized by a magnetic mirror, with its inevitable imperfections,
leading to nonideal efficiency. The main purpose of this article
is to quantitatively analyze the consequences of these
imperfections: clearly, they tend to deteriorate the performance
of the experimental setup; yet, for reasonable values of the
experimental parameters
\cite{ref:NeutronStorage1991,ref:RauchQZE}, a good test is still
clearly feasible with high efficiency. This will be shown in Sec.\
\ref{sec:QZENonIdeal}, where we will determine an \textit{optimum}
value $N_\text{opt}$ of the frequency of measurements: more
frequent measurements would simply deteriorate the overall
performance of the setup, masking the QZE\@. These conclusions are
of general validity: the presence of losses and imperfections
always leads to an optimal frequency, which is in general finite.
Our analysis will be extended and generalized in Sec.\
\ref{sec:genframe} to an arbitrary lossy quantum Zeno experiment,
and a convenient formula for $N_\text{opt}$ will be derived. We
summarize our results in Sec.\ \ref{sec:summa}\@.

\section{Neutron-Spin Test of the QZE with Ideal Mirrors}
\label{sec:QZEIdeal}
\begin{figure}[t]
\includegraphics[width=0.47\textwidth]{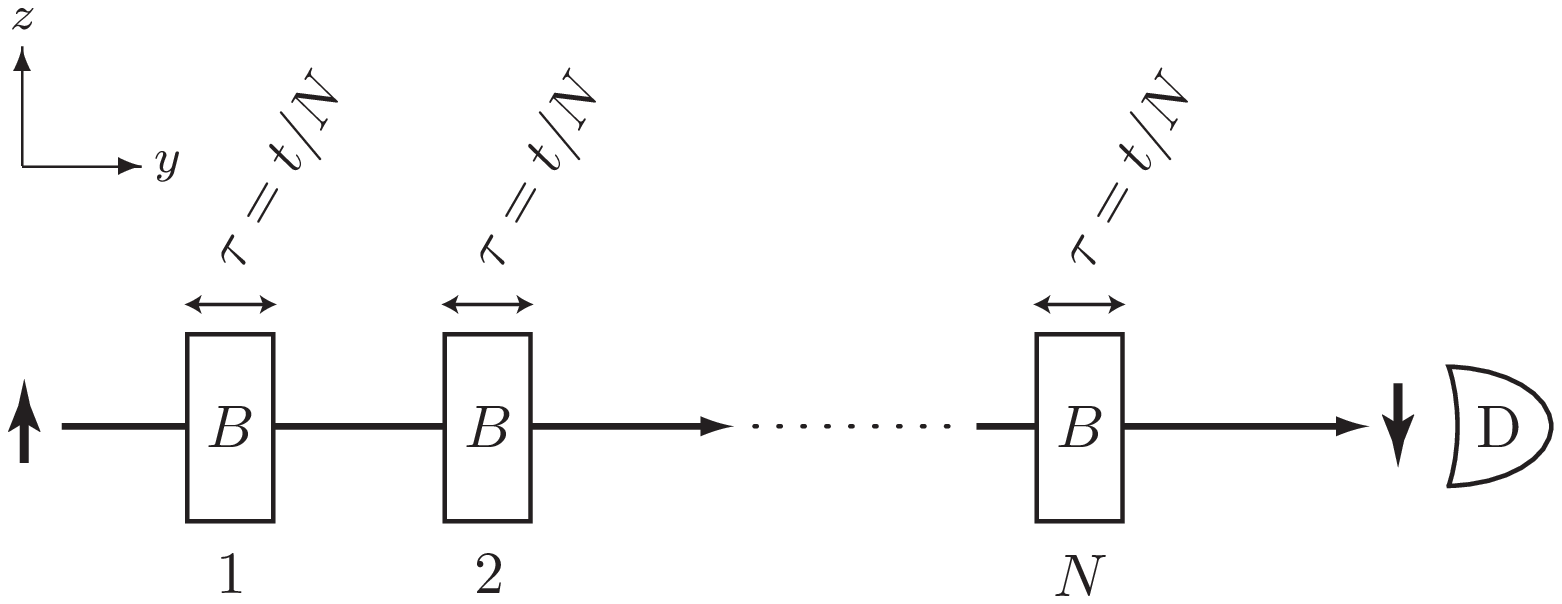}\\
\begin{picture}(0,0)
\put(12,83){\makebox(0,0){(a)}}
\end{picture}
\bigskip\\
\includegraphics[width=0.47\textwidth]{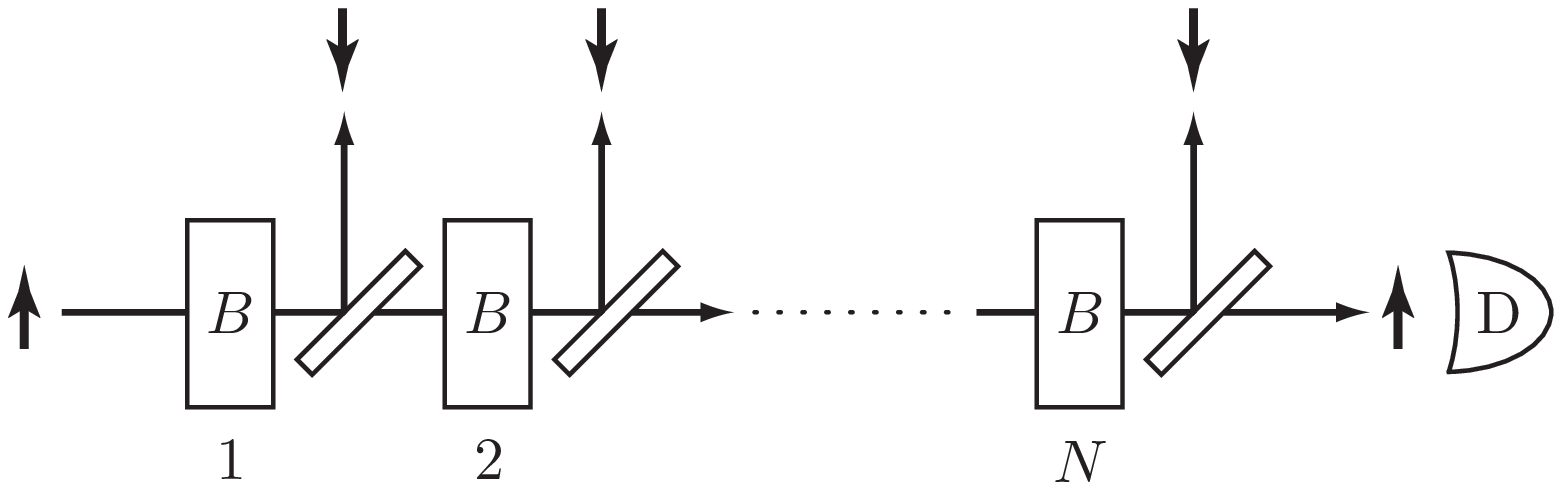}\\
\begin{picture}(0,0)
\put(12,74){\makebox(0,0){(b)}}
\end{picture}
\caption{(a) Basic setup for the neutron-spin test of QZE\@.
We set $t/\tau_\text{Z}=\pi/2$, so that $2\theta=\pi$. (b)
Neutron-spin test of QZE with ideal mirrors. }
\label{fig:TestIdeal}
\end{figure}

Let us first briefly review the original proposal of the neutron
spin test of the QZE \cite{ref:NeutronSpinTestQZE}. The basic
setup is shown in Fig.\ \ref{fig:TestIdeal}(a).
We prepare, equally spaced along the $y$ axis, $N$ identical
regions in each of which a static magnetic field $B$ is applied in
the $x$ direction. A neutron wave packet, whose initial spin is
oriented in the $z$ direction, travels along the $y$ axis and
undergoes a spin rotation at each interaction with the magnetic
field, according to the Hamiltonian
\begin{equation}
H=\mu B\sigma_x,
\label{eqn:simpleHam}
\end{equation}
$\mu$ being the neutron magnetic moment and $\sigma_i$ ($i=x,y,z$)
the Pauli matrices. The initial state of the incident neutron is
$\ket{S_0}=\ket{\uparrow}$ (spin up along the $z$ direction). The
final state, after crossing the $N$ regions with the magnetic
fields, reads
\begin{equation}
\ket{S(t)}
=e^{-iHt/\hbar}\ket{\uparrow} =\cos\frac{\mu
Bt}{\hbar}\ket{\uparrow} -i\sin\frac{\mu
Bt}{\hbar}\ket{\downarrow},
\end{equation}
where $t$ is the total time spent in the magnetic field and we
have ignored, for simplicity, the spatial degrees of freedom of
the neutron. By defining
\begin{equation}\label{eqn:BT}
\theta=\frac{\mu Bt}{\hbar}=\frac{t}{\tau_\text{Z}} ,
\end{equation}
where $\tau_\text{Z}\,(=\hbar\langle {\uparrow} |H^2|{\uparrow}
\rangle^{-1/2}$ in this case) is the so-called Zeno time and
$2\theta$ the classical precession angle of the spin, the survival
probability of the initial state $\ket{\uparrow}$ reads
\begin{equation}
P(\theta)=|\bra{\uparrow}e^{-iHt/\hbar}\ket{\uparrow}|^2=
\cos^2\!\theta .
\label{eqn:PNoMirrorgen}
\end{equation}
Notice that if $Bt$ is adjusted so as to satisfy
\begin{equation}\label{eqn:BT11}
\theta=\frac{\pi}{2},
\end{equation}
the spin is completely flipped
\begin{equation}
\ket{S(t)}=e^{-iHt/\hbar}\ket{\uparrow}=-i\ket{\downarrow}.
\end{equation}
In this case the survival probability of the initial state
$\ket{\uparrow}$ vanishes
\begin{equation}
P(\theta)= 0.
\label{eqn:PNoMirror}
\end{equation}
This situation, shown in Fig.\ \ref{fig:TestIdeal}(a), is that
usually considered in the literature. However, the whole analysis
that follows identically applies to the general case
(\ref{eqn:BT})--(\ref{eqn:PNoMirrorgen}).

Let us now check, at every step, whether the spin has remained in
the initial state $\ket{\uparrow}$ despite the spin rotation in
the $B$-field. To this end, we insert $N$ magnetic mirrors after
every $B$-region, as in Fig.\ \ref{fig:TestIdeal}(b). The incident
neutron undergoes $N$ ``spin-measurements'' until it reaches the
detector D\@. At each step, if the spin state remains up, the
neutron is transmitted through the mirror and keeps traveling
right, otherwise it is reflected out by the mirror. Detector D
counts those neutrons that have ``survived'' at each of these $N$
``measurements,'' so that the detection probability at D is
nothing but the survival probability of the initial state
$\ket{\uparrow}$.

As clarified in Refs.\ \cite{ref:reviewQZE}
and \cite{ref:NeutronSpinTestQZE}, the insertion of a mirror does
not represent a measurement of the spin state; it just constitutes
a generalized spectral decomposition (GSD) in Wigner's
sense \cite{Wigner}, namely a (unitary) physical process that
associates an ``external'' degree of freedom (whose role is played
here by the wave packet of the neutron) to different values of the
observable to be measured (the neutron spin): a frequent sequence
of GSD is sufficient for the occurrence of a QZE\@. In a magnetic
field, the spin state of the incident neutron is changed from the
initial one $\ket{\uparrow}$ to $e^{-iHt/N\hbar}\ket{\uparrow}$
and the neutron is then
\textit{decomposed} by the mirror into two branch waves: the
spin-up component going rightward and the spin-down one going
upward in Fig.\ \ref{fig:TestIdeal}(b). The state of the neutron
just after the first mirror is hence given by
\begin{equation}
\ket{\psi_1}
=\mathcal{T}e^{-iHt/N\hbar}\ket{\uparrow}\otimes\ket{t_1}
+\mathcal{R}e^{-iHt/N\hbar}\ket{\uparrow}\otimes\ket{r_1},
\label{eqn:MirrorActionIdeal}
\end{equation}
where the spectral decomposition with respect to the spin state is
expressed in terms of the projection operators
\begin{equation}
\mathcal{T}=\ket{\uparrow}\bra{\uparrow},\quad
\mathcal{R}=\ket{\downarrow}\bra{\downarrow}
\label{eqn:IdealProjectionOp}
\end{equation}
and $\ket{t_n}$ and $\ket{r_n}$ are the transmitted and reflected
wave packets after the $n$th mirror [and before the $(n+1)$th
magnetic field], representing the spatial degrees of freedom of
the neutron. Repeating these operations $N$ times, we obtain the
final state of the neutron
\begin{eqnarray}
\ket{\psi_N}
&=&(\mathcal{T}e^{-iHt/N\hbar})^N\ket{\uparrow}\otimes\ket{t_N}
\nonumber\\
&&{}+\sum_{n=1}^N\mathcal{R}e^{-iHt/N\hbar}
(\mathcal{T}e^{-iHt/N\hbar})^{n-1}
\ket{\uparrow}\otimes\ket{r_n},\nonumber\\
\label{eqn:FinalStateIdeal}
\end{eqnarray}
so that the probability for the neutron to be detected at detector
D, i.e., the survival probability of the initial spin state
$\ket{\uparrow}$, reads
\begin{eqnarray}
&&P^{(N)}(\theta)\nonumber\\
&&\quad=|\{\bra{\uparrow}\otimes\bra{t_N}\}\ket{\psi_N}|^2
=|\bra{\uparrow}(\mathcal{T}e^{-iHt/N\hbar})^N\ket{\uparrow}|^2
\nonumber\\
&&\quad=|\bra{\uparrow}e^{-iHt/N\hbar}\ket{\uparrow}|^{2N}
=\left(\cos\frac{\theta}{N}\right)^{2N},
\label{eqn:PIdeal}
\end{eqnarray}
where we have made use of Eq.\ (\ref{eqn:BT}) (within our
approximations, the total duration of the experiment is $t$, with
or without magnetic mirrors). Under the condition (\ref{eqn:BT11})
(and in general for $\theta<\pi/2$), this is nonvanishing for any
$N\ge2$ and is an increasing function of $N$\@. Frequent
``checks'' of the spin state slow down the evolution of the
initial state $\ket{\uparrow}$: the survival probability
$P^{(N)}(\theta)$ increases with the frequency of
``measurements.'' This is a QZE\@. Furthermore, in the limit of
infinite frequency,
\begin{equation}
\lim_{N\to\infty}P^{(N)}(\theta)=1\quad\text{($\theta$ fixed)},
\end{equation}
i.e., the spin is frozen and ceases to evolve, in agreement with
the theorem by Misra and Sudarshan \cite{ref:QZEMisraSudarshan}.

An experiment is at present being performed \cite{ref:RauchQZE} by
making use of a recently developed neutron storage
technique \cite{ref:NeutronStorage1991}. Neutrons with a
well-defined energy and in a given spin state are stored in a
$1\,\mathrm{m}$ long perfect crystal resonator. The neutrons, at
the given energy, satisfy the Bragg reflection condition and
bounce back and forth between the two slabs at both ends of the
silicon crystal. (At present, neutrons can be reflected a few
thousands times with small
losses \cite{ref:NeutronStorage1991,ref:RauchQZE}.) In the central
part of the resonator, a spin-rotating RF field will be applied,
playing the role of the magnetic field in
Fig.\ \ref{fig:TestIdeal}\@.

The Zeno effect can be obtained as follows. A neutron whose
wavelength satisfies the Bragg condition is reflected back by the
crystal. However, if a magnetic field is applied at one of the
crystal slabs, yielding different potentials for different spin
states of the neutron, the neutrons are selected according to
their spin state: if, say, a spin-up neutron satisfies the Bragg
condition at a plate, the neutron is reflected back and kept
inside the resonator; if, on the other hand, the spin is flipped
by the spin-rotating RF field, its wavelength does not meet the
Bragg condition and the neutron is transmitted out of the
resonator. The crystal plates with the magnetic fields play
therefore the role of the ``magnetic mirrors'' in
Fig.\ \ref{fig:TestIdeal}(b), performing the GSDs. Hence, in this
experimental setup, the probability for the neutron to remain in
the storage apparatus is the survival probability of the initial
spin state.

It should be clear by now that it is of primary importance to
analyze the effect of losses and imperfections, in order to
understand whether the experiment is still meaningful in a
realistic situation. Notice that the number $N$ of traverses and
interactions should be very large, in order to get a good
manifestation of the QZE\@. This, on the other hand, entails a
dramatic (exponential) propagation of ``errors.'' This will be
investigated in the following two sections.

\section{Neutron-Spin Test of the QZE with Non-Ideal Mirrors}
\label{sec:QZENonIdeal}
\begin{figure}[t]
\begin{tabular}{c@{\quad}c}
\includegraphics[width=0.22\textwidth]{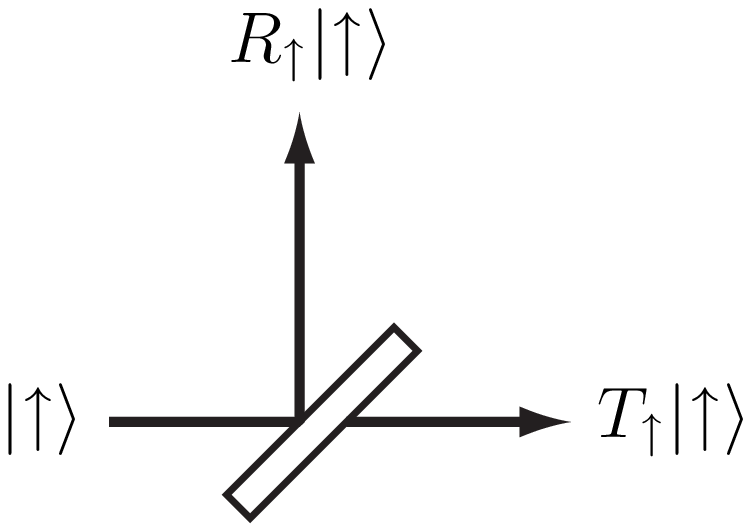}&
\includegraphics[width=0.22\textwidth]{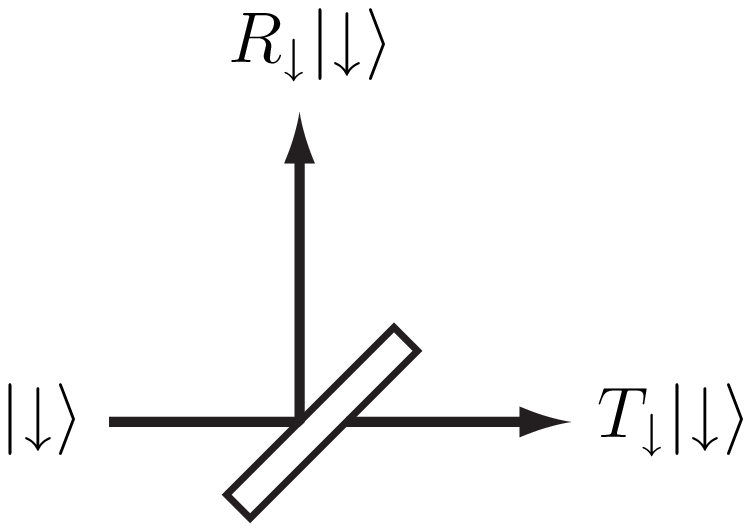}\\
(a)&(b)
\end{tabular}
\caption{Transmission and reflection coefficients for (a) a
spin-up neutron and (b) a spin-down neutron.}
\label{fig:NonIdealMirror}
\end{figure}

Losses are unavoidable in real experiments and must be duly taken
into account. A magnetic mirror, for example, is not ideal, as
tacitly assumed in the previous section. It has a nonvanishing
probability of failing to correctly decompose the spin state.
Assume that the magnetic mirror has transmission
$T_{\uparrow(\downarrow)}$ and reflection
$R_{\uparrow(\downarrow)}$ coefficients for a spin-up (spin-down)
neutron (Fig.\ \ref{fig:NonIdealMirror}). (They are in general
complex valued and constrained by
$|T_{\uparrow(\downarrow)}|^2+|R_{\uparrow(\downarrow)}|^2=1$.)
We assumed in the previous section that
$|T_\uparrow|=|R_\downarrow|=1$ and $R_\uparrow=T_\downarrow=0$,
but this is not the case for actual magnetic mirrors. So the
question arises as to whether (and to which extent) it is possible
to observe the QZE with non-ideal mirrors. In other words, whether
the QZE still takes place if the ``measurements'' (i.e., the
spectral decompositions) are \textit{imperfect}.

At the $n$th (non-ideal) mirror, the spin-up component of a
neutron, $\ket{\uparrow}\otimes\ket{t_{n-1}}$, is split into two
waves
\begin{subequations}
\label{eqn:NonIdealSpectralDecomposition}
\begin{equation}
\ket{\uparrow}\otimes\ket{t_{n-1}}
\to\ket{\uparrow}\otimes\Bigl(
T_\uparrow\ket{t_n}+R_\uparrow\ket{r_n}
\Bigr)
\end{equation}
and a similar expression holds for the spin-down component
\begin{equation}
\ket{\downarrow}\otimes\ket{t_{n-1}}
\to\ket{\downarrow}\otimes\Bigl(
T_\downarrow\ket{t_n}+R_\downarrow\ket{r_n}
\Bigr).
\end{equation}
\end{subequations}
(No spin-flip is assumed to occur at the magnetic mirror. The most
general case, where such spin-flips take place, is investigated in
the Appendix.) The right arrows in
Eqs.\ (\ref{eqn:NonIdealSpectralDecomposition}) and in the
following stand for the (unitary) physical processes that are
responsible for the spectral decomposition. Hence for a neutron
in a general spin state $\ket{S}$, the magnetic mirror provokes
the following spectral decomposition
\begin{eqnarray}
\ket{S}\otimes\ket{t_{n-1}}
&\equiv&\Bigl(
c_\uparrow\ket{\uparrow}+c_\downarrow\ket{\downarrow}
\Bigr)\otimes\ket{t_{n-1}}\nonumber\\
&\to&\Bigl(
c_\uparrow T_\uparrow\ket{\uparrow}
+c_\downarrow T_\downarrow\ket{\downarrow}
\Bigr)\otimes\ket{t_n}\nonumber\\
&&{}+\Bigl(
c_\uparrow R_\uparrow\ket{\uparrow}
+c_\downarrow R_\downarrow\ket{\downarrow}
\Bigr)\otimes\ket{r_n}\nonumber\\
&=&\tilde{\mathcal{T}}\ket{S}\otimes\ket{t_n}
+\tilde{\mathcal{R}}\ket{S}\otimes\ket{r_n},
\label{eq:SD}
\end{eqnarray}
where the operators
\begin{equation}
\tilde{\mathcal{T}}
=\ket{\uparrow}T_\uparrow\bra{\uparrow}
+\ket{\downarrow}T_\downarrow\bra{\downarrow},\quad
\tilde{\mathcal{R}}
=\ket{\uparrow}R_\uparrow\bra{\uparrow}
+\ket{\downarrow}R_\downarrow\bra{\downarrow}
\label{eqn:PseudoProjectionOp}
\end{equation}
incorporate the effects due to the imperfections of the mirror.
These operators $\tilde{\mathcal{T}}$ and $\tilde{\mathcal{R}}$,
even though they are no longer projection operators, play the same
role as the projection operators $\mathcal{T}$ and $\mathcal{R}$
in the ideal case
(\ref{eqn:MirrorActionIdeal})--(\ref{eqn:IdealProjectionOp}). The
final state of the neutron after the final ($N$th) magnetic mirror
is given by
\begin{eqnarray}
\ket{\tilde{\psi}_N}
&=&(\tilde{\mathcal{T}}e^{-iHt/N\hbar})^N\ket{\uparrow}\otimes
\ket{t_N}\nonumber\\
&&{}+\sum_{n=1}^N\tilde{\mathcal{R}}e^{-iHt/N\hbar}
(\tilde{\mathcal{T}}e^{-iHt/N\hbar})^{n-1}\ket{\uparrow}\otimes
\ket{r_n}\nonumber\\
\label{eqn:FinalStateNonIdeal}
\end{eqnarray}
and the probability for the neutron to be detected at detector D
reads
\begin{eqnarray}
\tilde{P}^{(N)}(\theta)
&=&\|\bracket{t_N}{\tilde{\psi}_N}\|^2\nonumber\\
&=&\tr[(\tilde{\mathcal{T}}e^{-iHt/N\hbar})^N\rho_0
(e^{iHt/N\hbar}\tilde{\mathcal{T}}^\dag)^N],\quad
\label{eqn:PNonIdealDef}
\end{eqnarray}
where $\rho_0=\ket{\uparrow}\bra{\uparrow}$ is the initial density
operator of the neutron spin. [The spin state observed at the
detector is not necessarily $\ket{\uparrow}$; it is the
probability (\ref{eqn:PNonIdealDef}) that one measures in the
actual experiment.]

Let us evaluate the probability (\ref{eqn:PNonIdealDef}).
The eigenvalues $\xi_\pm(N)$ of the operator
\begin{eqnarray}
&&\tilde{\mathcal{T}}e^{-iHt/N\hbar}\nonumber\\
&&\quad=\frac{1}{2}(T_\uparrow+T_\downarrow)\cos\frac{\theta}{N}
-\sigma_x\frac{i}{2}(T_\uparrow+T_\downarrow)\sin\frac{\theta}{N}
\nonumber\\
&&\quad\phantom{=}
{}+\sigma_y\frac{1}{2}(T_\uparrow-T_\downarrow)\sin\frac{\theta}{N}
+\sigma_z\frac{1}{2}(T_\uparrow-T_\downarrow)\cos\frac{\theta}{N}
\nonumber\\
\label{eqn:TransferMatrix}
\end{eqnarray}
are given by
\begin{eqnarray}
\xi_\pm(N)
&=&\frac{1}{2}\,\Biggl[
(T_\uparrow+T_\downarrow)\cos\frac{\theta}{N}\nonumber\\
&&\phantom{\frac{1}{2}\,\Biggl[} {}\pm\sqrt{
(T_\uparrow+T_\downarrow)^2\cos^2\!\frac{\theta}{N} -4T_\uparrow
T_\downarrow }
\Biggr].
\nonumber \\
\label{eqn:Eigenvalues}
\end{eqnarray}
[The eigenvalues $\xi_\pm(N)$ will henceforth be written
$\xi_\pm$, unless confusion arises.] By rewriting the operator
(\ref{eqn:TransferMatrix}) as
\begin{equation}
\tilde{\mathcal{T}}e^{-iHt/N\hbar}
=\frac{1}{2}(\xi_++\xi_-) +\frac{1}{2}(\xi_+-\xi_-)\sigma_n,
\end{equation}
where $\sigma_n=\bm{n}\cdot\bm{\sigma}$, $\bm{n}$ being a
complex-valued vector satisfying $\bm{n}^2=n_x^2+n_y^2+n_z^2=1$,
we readily obtain
\begin{equation}
(\tilde{\mathcal{T}}e^{-iHt/N\hbar})^N=
\frac{1}{2}(\xi_+^N+\xi_-^N)
+\frac{1}{2}(\xi_+^N-\xi_-^N)\sigma_n.
\end{equation}
A series of elementary calculations yields the following exact
expression for the probability
\begin{eqnarray}
&&\tilde{P}^{(N)}(\theta)\nonumber\\
&&\quad=\left| A(N)-B(N)T_\downarrow\cos\frac{\theta}{N}
\right|^2+\left|
B(N)T_\downarrow\sin\frac{\theta}{N}
\right|^2, \nonumber\\
\label{eqn:PNonIdeal}
\end{eqnarray}
with
\begin{subequations}
\begin{equation}
A(N)=\frac{\xi_+^{N+1}(N)-\xi_-^{N+1}(N)}{\xi_+(N)-\xi_-(N)},
\label{eqn:A}
\end{equation}
\begin{equation}
B(N)=\frac{\xi_+^N(N)-\xi_-^N(N)}{\xi_+(N)-\xi_-(N)}.
\label{eqn:B}
\end{equation}
\end{subequations}

We are now in a position to see whether it is possible to observe
the QZE with non-ideal mirrors. In order to analyze its
$N$-dependence, let us expand the probability
(\ref{eqn:PNonIdeal}) as a function of
$|T_\downarrow/T_\uparrow|\ll1$. (In the experiment
\cite{ref:NeutronStorage1991},
$|T_\downarrow/T_\uparrow|^2\lesssim10^{-4}$.) For any $N\ge2$,
the eigenvalues in Eq.\ (\ref{eqn:Eigenvalues}) are expanded as
\begin{subequations}
\begin{equation}
\xi_+\simeq T_\uparrow\cos\frac{\theta}{N}\left[
1-\frac{T_\downarrow}{T_\uparrow}\tan^2\!\frac{\theta}{N}
+O(T_\downarrow^2/T_\uparrow^2)
\right],
\end{equation}
\begin{equation}
\xi_-\simeq\xi_+\left[
\frac{T_\downarrow}{T_\uparrow}
\left(1+\tan^2\!\frac{\theta}{N}\right)
+O(T_\downarrow^2/T_\uparrow^2)
\right],
\end{equation}
\end{subequations}
from which one obtains
\begin{eqnarray}
A(N)
&=&\xi_+^N\,\biggl[
1+\frac{\xi_-}{\xi_+}+\cdots+\left(\frac{\xi_-}{\xi_+}\right)^N
\biggr]\nonumber\\
&\simeq&\left(T_\uparrow\cos\frac{\theta}{N}\right)^N\nonumber\\
&&{}\times
\left[
1-\frac{T_\downarrow}{T_\uparrow}\left(
(N-1)\tan^2\!\frac{\theta}{N}-1
\right)+\cdots
\right]\nonumber\\
\end{eqnarray}
and a similar expansion holds for $B(N)$. We thus easily obtain an
approximate expression for the probability (\ref{eqn:PNonIdeal})
\begin{eqnarray}
\tilde{P}^{(N)}(\theta)
&\simeq&|T_\uparrow|^{2N}
\left(\cos\frac{\theta}{N}\right)^{2N}\nonumber\\
&&{}\times\left[
1-2\Re\!\left(\frac{T_\downarrow}{T_\uparrow}\right)
(N-1)\tan^2\!\frac{\theta}{N} +\cdots
\right], \nonumber\\
\label{eqn:PExpansion}
\end{eqnarray}
valid for $N\ge2$. [For $N=1$,
$\tilde{P}^{(N)}(\theta)=\sin^2\!\theta\,|T_\downarrow|^2
+\cos^2\!\theta\,|T_\uparrow|^2$ exactly.] It is clear from
formula (\ref{eqn:PExpansion}) that the probability
$\tilde{P}^{(N)}(\theta)$ is well approximated by
\begin{figure*}
\begin{tabular}{c@{\qquad}c}
\medskip
\includegraphics[height=0.31\textwidth]{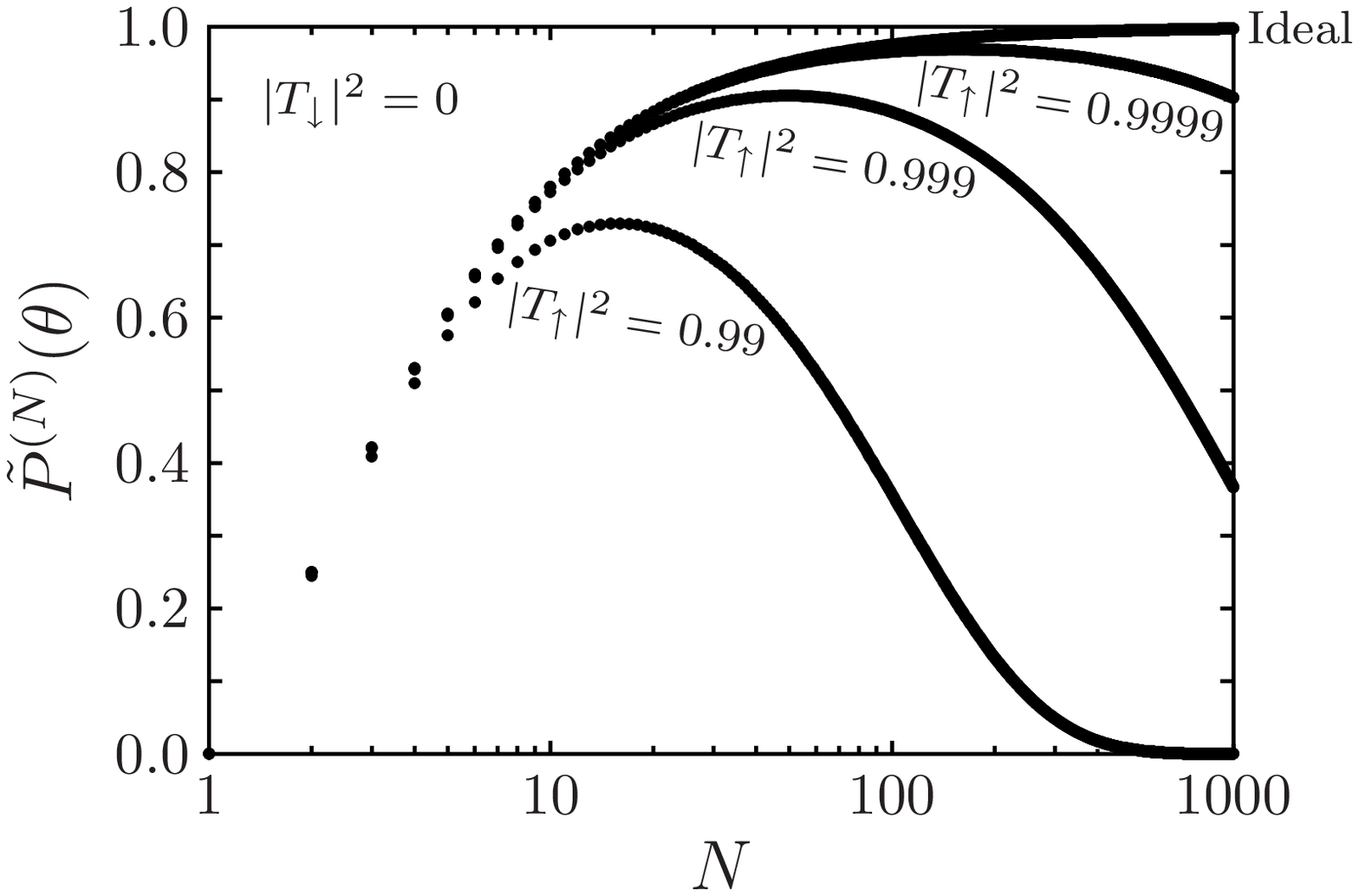}&
\includegraphics[height=0.31\textwidth]{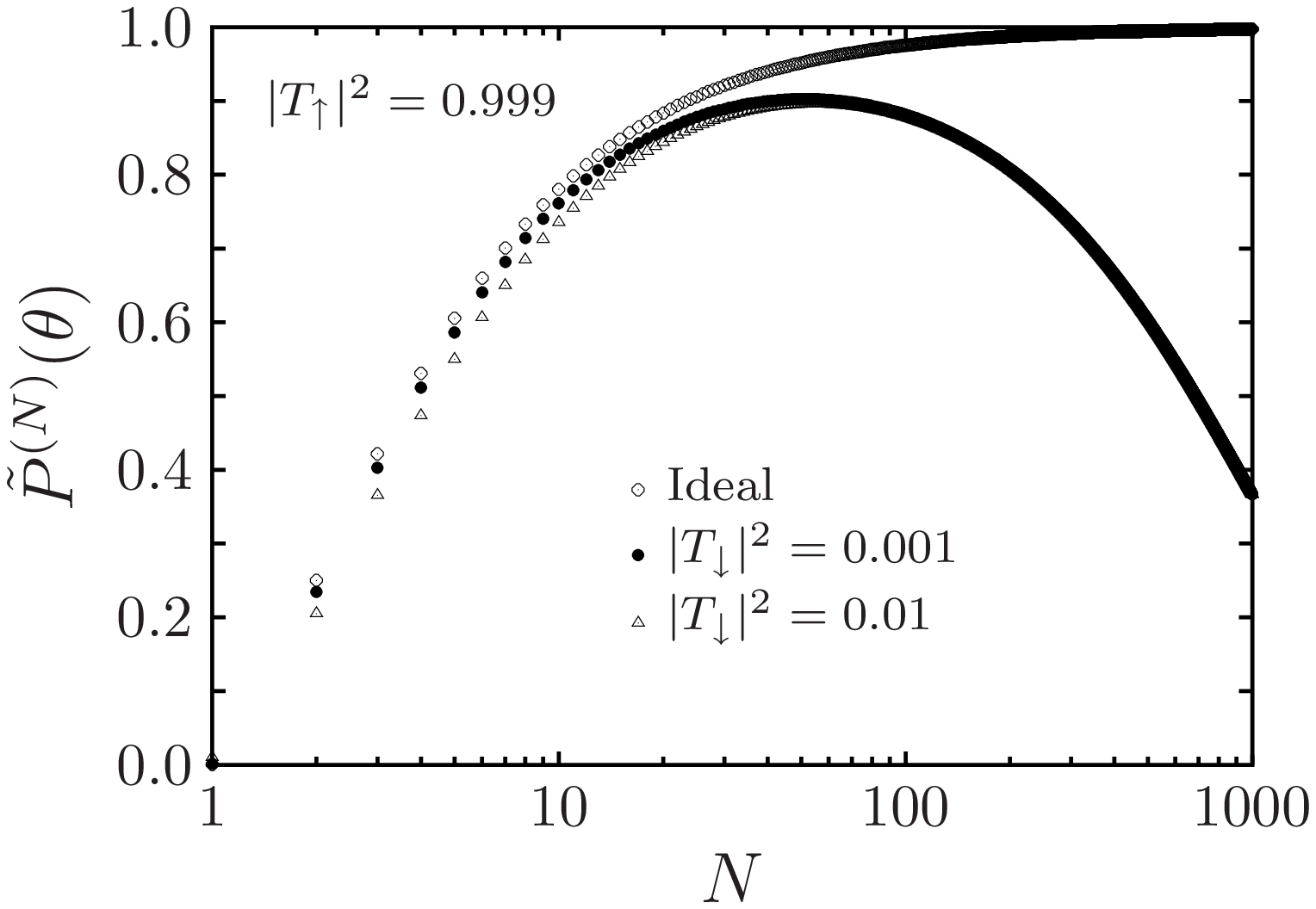}\\
(a)&(b)
\end{tabular}
\caption{(a) $T_\uparrow$-dependence and (b)
$T_\downarrow$-dependence of the probability
$\tilde{P}^{(N)}(\theta)$ in Eq.\ (\ref{eqn:PNonIdeal}). In both
figures, $\arg T_\uparrow=\arg T_\downarrow=0$.}
\label{fig:PNonIdeal}
\end{figure*}
\begin{equation}
\tilde{P}^{(N)}(\theta)
\simeq|T_\uparrow|^{2N}\left(\cos\frac{\theta}{N}\right)^{2N}.
\label{eqn:PApp}
\end{equation}
This shows that neither the transmission coefficient
$T_\downarrow$ for a spin-down neutron, nor the phases of
$T_\uparrow$ and $T_\downarrow$ bear any important influence on
the probability $\tilde{P}^{(N)}(\theta)$; the only relevant
quantity is the transmission probability $|T_\uparrow|^2$. Since
$|T_\uparrow|^2 \simeq 1$, for $N$ not too large the factor
$|T_\uparrow|^{2N}$ is almost unity and the probability
$\tilde{P}^{(N)}(\theta)$ behaves like
\begin{equation}
\tilde{P}^{(N)}(\theta)
\simeq\left(\cos\frac{\theta}{N}\right)^{2N}\quad
\text{($N$ not too large)}.
\end{equation}
This is the same as the survival probability with ideal mirrors
given in Eq.\ (\ref{eqn:PIdeal}), and is an increasing function of
$N$. However, for larger $N$, the factor $[\cos(\theta/N)]^{2N}$
is almost unity, and the probability behaves like
\begin{equation}
\tilde{P}^{(N)}(\theta)\simeq|T_\uparrow|^{2N}\quad \text{(larger
$N$)} ,
\end{equation}
decreasing exponentially to zero as $N\to\infty$: as the number of
mirrors, $N$, is increased, the mirror imperfections
($|T_\uparrow|^{2N} <1$) dominate over the increasing factor
$[\cos(\theta/N)]^{2N}$, suppressing the QZE for very large $N$.
(Clearly, the meaning of ``large'' $N$ in the two preceding
equations must be precisely defined. This will be done in the
following.)

There must be therefore an optimal number of mirrors,
$N_\text{opt}$, in order to observe the QZE if the losses in the
``measurement'' processes (spectral decompositions) are taken into
account. In Fig.\ \ref{fig:PNonIdeal}, the probability
$\tilde{P}^{(N)}(\theta)$ computed according to the exact
expression (\ref{eqn:PNonIdeal}) is plotted as a function of $N$
for a few values of the transmission coefficients $T_\uparrow$ and
$T_\downarrow$. The figures corroborate the previous discussion.
The QZE can be observed even with non-ideal mirrors, if $N$ is not
too large, namely if one does not ``check'' the system's state too
frequently: this is good news from an experimental point of view,
since one need not and should not attempt to indefinitely increase
the number of mirrors (or reflections in the neutron resonator
experiment) in order to achieve an optimal QZE\@. Notice also that
the probability $\tilde{P}^{(N)}(\theta)$ significantly depends on
$T_\uparrow$, but displays almost no dependence on $T_\downarrow$.

It is possible to estimate the optimal number of ``measurements,''
$N_\text{opt}$, yielding the maximum probability
$\tilde{P}^{(N_\text{opt})}(\theta)$. This can be done from the
approximate formula (\ref{eqn:PApp}) as follows. For actual
magnetic mirrors, $|T_\uparrow|^2$ is almost unity (a reasonable
value of $1-|T_\uparrow|^2$ is of order
$10^{-4}$ \cite{ref:NeutronStorage1991}) and $N_\text{opt}$ is
expected to be large. The maximum of the function
$f(x)=a^x\cos^x(2 \theta/x)$, with $a \lesssim 1$, is given by one
of the solutions of the equation $a\cos(2\theta/x)%
=\exp[-(2\theta/x)\tan(2 \theta/x)]$ and is approximately
$x_\text{opt}\simeq2 \theta/\sqrt{\ln a^{-2}}$. Applying this
result to the probability (\ref{eqn:PApp}) one obtains
\begin{equation}
N_\text{opt}
\simeq \left[\frac{\theta}{\sqrt{1-|T_\uparrow|^2}}\right]
\quad(|T_\uparrow|^2\simeq1),
\label{eqn:NPeak}
\end{equation}
where $[x]$ is the closest integer to $x$.
The maximum is then readily evaluated
\begin{subequations}
\label{eqn:PvsN}
\begin{alignat}{2}
\tilde{P}^{(N_\text{opt})}(\theta)
&\simeq1-\frac{2\theta^2}{N_\text{opt}}&\quad
(N_\text{opt}\gg1)\\
&\simeq1-2\theta\sqrt{1-|T_\uparrow|^2}&\quad
(|T_\uparrow|^2\simeq1).
\end{alignat}
\end{subequations}
Some values of $N_\text{opt}$ and
$\tilde{P}^{(N_\text{opt})}(\theta)$ estimated with
Eqs.\ (\ref{eqn:NPeak}) and (\ref{eqn:PvsN}), respectively, are
listed in Table \ref{tab:NPeak} for some $|T_\uparrow|^2$. The
agreement with the numerical results shown in
Fig.\ \ref{fig:PNonIdeal}, based on the exact
formula (\ref{eqn:PNonIdeal}), is excellent [except for
$|T_\uparrow|^2=0.99$, where $\tilde{P}^{(N_\text{opt})}(\theta)$
differs by about $5\%$].
\begin{table}[t]
\catcode`?=\active \def?{\phantom{0}}
\caption{$N_\text{opt}$ from
Eq.\ (\ref{eqn:NPeak}) and $\tilde{P}^{(N_\text{opt})}(\theta)$
from Eq.\ (\ref{eqn:PvsN}) versus $|T_\uparrow|^2$. The exact
values, obtained from Eq.\ (\ref{eqn:PNonIdeal}) with
$|T_\downarrow|^2=0$, are indicated in parentheses.}
\label{tab:NPeak}
\begin{ruledtabular}
\begin{tabular}{ccc}
$|T_\uparrow|^2$&$N_\text{opt}$ &
$\tilde{P}^{(N_\text{opt})}(\theta)$\\\hline
$0.99??$&$?16\,?(16) $&$0.69\,(0.73)$\\
$0.999?$&$?50\,?(50) $&$0.90\,(0.91)$\\
$0.9999$&$157\,(157) $&$0.97\,(0.97)$
\end{tabular}
\end{ruledtabular}
\end{table}

Notice that for
$1-|T_\uparrow|^2\sim10^{-4}$ \cite{ref:NeutronStorage1991}, the
estimated optimal number is $N_\text{opt}=157$, which is much
smaller than the so-far achievable number of traverses
$N_\text{max}\sim4000$ in the
experiment \cite{ref:NeutronStorage1991,ref:RauchQZE}; yet the
survival probability
$\tilde{P}^{(N_\text{opt})}(\theta)\simeq0.97$ is already very
close to unity. This estimate shows that a good test of the QZE
can be performed in this case.

Of course, actual experiments suffer from other losses than those
considered here. However, such additional losses can be taken into
account (to a large extent), by duly renormalizing the
transmission probability $|T_\uparrow|^2$. We therefore expect
that the present analysis essentially maintains its validity. For
example, if the maximum number of traverses in a neutron-spin test
of the QZE is of order $N_\text{max}\simeq4000$, one can roughly
estimate that $1-|T_\uparrow|^2\sim$ losses $ \simeq 1/4000$. This
yields $N_\text{opt}\simeq 99$ and
$\tilde{P}^{(N_\text{opt})}(\theta)\simeq0.95$, a very reasonable
value.

\section{QZE with Non-Ideal Measurements: General Framework}
\label{sec:genframe}
It is possible to extend the conclusions of the preceding section
to a broader framework, by making use of the well-known
characteristics of the QZE (short-time behavior of the evolved
wave function) and of some sensible assumptions regarding the
GSD\@. Assume that $N$ is large and the losses small, so that the
quantum Zeno survival probability be given by an expression of the
type (\ref{eqn:PExpansion})--(\ref{eqn:PApp}),
\begin{equation}
\tilde{P}^{(N)}(\theta) \simeq [L(t_1/N)]^N [p(t_2/N)]^N\quad
(t_1+t_2 = t= \tau_\text{Z} \theta),
\label{eqn:Pgen}
\end{equation}
where the factor $L$ represents losses (due to imperfect
transmission, measurements, and so on), while $p$ is the survival
probability of the quantum system in its initial state. We require
that
\begin{equation}
0 \leq L(t), p(t) \leq 1.
\label{eqn:PLreq}
\end{equation}
Equations (\ref{eqn:Pgen})--(\ref{eqn:PLreq}) describe the Zeno
survival probability in an experiment in which a quantum evolution
followed by a lossy spectral decomposition is repeated $N$ times.
In short, the system spends a time $t_2$ evolving under the action
of a given Hamiltonian $H$ and a time $t_1$ in GSDs. (We notice
that $t_2$ plays the same role as $t$ of the previous section,
where the GSD time $t_1$ was neglected.) We will write
\begin{equation}
t_j=\alpha_jt,\ \alpha_j>0\ (j=1,2),\quad\alpha_1+\alpha_2=1.
\label{eqn:Tjj}
\end{equation}

The quantum mechanical survival probability has the following
short-time expansion \cite{ref:reviewQZE}
\andy{survp}
\begin{equation}
p(t)\sim1-\frac{t^2}{\tau_\text{Z}^2}\quad (t < \tau_\text{Z}),
\label{eqn:survp}
\end{equation}
where $\tau_\text{Z}$ is the Zeno time. Notice that in general
(and in particular for {\em bona fide} unstable systems) the above
equation is valid on a (much) {\em shorter} timescale than
$\tau_\text{Z}$, but this will not be discussed here: see
\cite{ref:Antoniou} and the last paper in \cite{ref:IQZE}.

We assume in general that
\andy{Lprop}
\begin{equation}
L(t)\sim a+bt+ct^2,\quad0\le a\le1\quad(\text{small } t).
\label{eqn:Lprop}
\end{equation}
When $a=1$, the GSD is very effective and losses appear on a
timescale of order $|b|^{-1}$. By contrast, when $a<1$, losses are
``instantaneous'' and have serious consequences on a realistic
test of the QZE\@. (Notice that the above formula includes the
case in which $L$ is independent of $t$, when $b=c=0$.)

The strategy is to maximize $\ln\tilde{P}^{(N)}(\theta)$ in
Eq.\ (\ref{eqn:Pgen}) as a function of $N$, at fixed $t_1$ and
$t_2$. We get
\andy{PNmax}
\barr
\frac{d}{dN} \ln\tilde{P}^{(N)}(\theta) & = & \ln L(t_1/N) + \ln
p(t_2/N) \nonumber \\
& &{}-\frac{t_1 L'(t_1/N)}{NL(t_1/N)} -\frac{t_2
p'(t_2/N)}{Np(t_2/N)}=0, \nonumber \\
\label{eqn:PNmax}
\earr
where the prime denotes derivative with respect to the whole
argument. By expanding for large $N$, according to
Eqs.\ (\ref{eqn:survp}) and (\ref{eqn:Lprop}), this yields
\andy{freqmeas}
\barr
\tau_\text{opt}^{-1} \equiv \frac{N_\text{opt}}{t} \simeq
\frac{\alpha_2}{\tau_\text{Z}\sqrt{\ln a^{-1}}}
\sqrt{1-\tau_\text{Z}^2\left(\frac{\alpha_1}{\alpha_2}\right)^2
\left(\frac{c}{a}-\frac{b^2}{2a^2}\right)}.
\nonumber \\
\label{eqn:freqmeas}
\earr
Plugging this result into (\ref{eqn:survp}), (\ref{eqn:Lprop}),
and (\ref{eqn:Pgen}), we obtain
\andy{eqn:PNlossy}
\barr
\lefteqn{\tilde{P}^{(N_\text{opt})}(\theta)}\nonumber\\
&\sim&\left[a+b\frac{t_1}{N_\text{opt}}+c
\left(\frac{t_1}{N_\text{opt}}\right)^2
\right]^{N_\text{opt}}\!
\left[1-\left(\frac{t_2}{\tau_\text{Z}N_\text{opt}}\right)^2
\right]^{N_\text{opt}} \nonumber \\
&\simeq&a^{N_\text{opt}}
\exp\!\left(\frac{b}{a}t_1 +\frac{c}{a} \frac{t_1^2}{N_\text{opt}}
-\frac{1}{2} \frac{b^2}{a^2} \frac{t_1^2}{N_\text{opt}}
-\frac{t_2^2}{\tau_\text{Z}^2N_\text{opt}} \right) \nonumber \\
&\simeq&a^{2N_{\text{opt}}}
\exp\!\left(\frac{b}{a}t_1 \right),
\label{eqn:PNlossy}
\earr
where we used (\ref{eqn:freqmeas}) in the last equality. The
factor $a^{2N_{\text{opt}}}$ is due to the two (almost equal)
terms $L$ and $p$ in (\ref{eqn:Pgen}), each contributing
$a^{N_{\text{opt}}}$. Equations (\ref{eqn:freqmeas}) and
(\ref{eqn:PNlossy}) are the main results of this section and
express the optimal frequency of GSDs, $\tau_\text{opt}^{-1}$, and
the maximal survival probability
$\tilde{P}^{(N_\text{opt})}(\theta)$ as a function of the
parameters characterizing the system and the apparatus.

Let us look at some particular cases.
If $a\to1$ (and $\forall
b,c$), corresponding to (almost) lossless GSDs,
$\tau_\text{opt}\to0$ and one gets the usual QZE, with no
limitations on the frequency of GSDs: infinitely frequent GSDs
slow down the evolution away from the initial quantum state.
However, due to the presence of losses, the survival probability
is not unity, even in the limit of infinitely frequent GSDs:
\andy{eqn:PNlossylim}
\beq
\tilde{P}^{(N_\text{opt})}(\theta)=\tilde{P}^{(\infty)}(\theta)
= \exp \left(-|b|t_1 \right) \quad (a \to 1),
\label{eqn:PNlossylim}
\eeq
where we took into account the fact that $b<0$ due to
(\ref{eqn:PLreq}) and $a=1$. This result is intuitively clear: due
to the presence of linear losses in $t$ in (\ref{eqn:Lprop}), one
cannot hope that the Zeno mechanism can work better than
(\ref{eqn:PNlossylim}). It is worth noticing that there are
analogies between this approach and interesting work by Berry and
Klein on twisted stacks of light polarizers \cite{BerryKlein}. It
should be emphasized that the practical limits one has to face in
the case of very frequent ``pulsed'' measurements ($N$ large) are
encompassed when one considers ``continuous'' measurement
processes, due to a Hamiltonian interaction with an external
system playing the role of apparatus. This is relevant in the
light of the physical equivalence between the ``pulsed'' and
``continuous'' formulations of the QZE \cite{Schulman98}.

If, on the other hand, $a \lesssim 1$, corresponding to
instantaneous losses, occurring on a GSD timescale (that we assume
to be much shorter than any other timescale: $t_1 \ll t_2\simeq
t$, or $\alpha_1 \ll\alpha_2\simeq 1$), Eq.\ (\ref{eqn:freqmeas})
yields
\andy{eqn:freqmeaslossy}
\beq
N_\text{opt}\simeq\frac{t}{\tau_\text{Z}\sqrt{\ln a^{-1}}} \simeq
\frac{t}{\tau_\text{Z}\sqrt{1-a}}.
\label{eqn:freqmeaslossy}
\eeq
This is the case considered in the previous section: if one
recalls the definition of $\theta$ in (\ref{eqn:BT}) and
identifies $a=|T_\uparrow|^2$, one recovers (\ref{eqn:NPeak}). In
this case the survival probability (\ref{eqn:PNlossy}) reduces to
(\ref{eqn:PvsN}).

Equations (\ref{eqn:freqmeas}) and (\ref{eqn:PNlossy}) enable one
to look at the ``lossy'' Zeno phenomenon from a more general
perspective. Clearly, in \emph{any} physical situation, the
optimal frequency (\ref{eqn:freqmeas}) to obtain a QZE is smaller
than $\infty$ and the optimal survival probability
(\ref{eqn:PNlossy}) is smaller than 1.

\section{Summary}
\label{sec:summa}
We have discussed a neutron-spin experimental test of the QZE from
a practical point of view, taking account of the inevitable
imperfection in the GSD at the magnetic mirror. We endeavored to
clarify that losses are important, but do not make an experimental
test of the QZE unrealistic. This is probably somewhat at variance
with expectation, for losses {\em exponentially} propagate in a
Zeno setup, involving $N$ repetitions of one and the same GSD\@.
However, we have seen that, if duly taken into account, the
disruptive effect of losses can be controlled and an interesting
test is still feasible for rather large values of $N$\@. This is a
positive conclusion, from an experimental perspective. Our
conclusions are of general validity for any practical test of the
QZE\@.

\begin{acknowledgments}
The authors acknowledge fruitful discussions and a useful exchange
of ideas with I.\ Ohba. K.Y.\ thanks L.\ Accardi and K.\ Imafuku
for enlightening comments and discussions. This work is partly
supported by Grants-in-Aid for Scientific Research (C) from the
Japan Society for the Promotion of Science (No.~14540280) and
Priority Areas Research (B) from the Ministry of Education,
Culture, Sports, Science and Technology, Japan (No.~13135221), by
a Waseda University Grant for Special Research Projects
(No.~2002A-567), and by the bilateral Italian-Japanese project
15C1 on ``Quantum Information and Computation'' of the Italian
Ministry for Foreign Affairs.
\end{acknowledgments}

\appendix*
\section{Spin-Flip effects at the Magnetic Mirrors}
In practice, one cannot exclude the possibility that a spin-flip
occurs at the magnetic mirrors. This effect introduces additional
mistakes and was neglected in Sec.\ \ref{sec:QZENonIdeal}\@. In
this Appendix, we take it into account and clarify its role in the
QZE\@.

The effects of the $n$th magnetic mirror on a spin-up and a
spin-down neutron read
\begin{eqnarray}
\ket{\uparrow}\otimes\ket{t_{n-1}}
&\to&\Bigl(
T_{\uparrow\uparrow}\ket{\uparrow}
+T_{\downarrow\uparrow}\ket{\downarrow}
\Bigr)\otimes\ket{t_n}\nonumber\\
&&{}+\Bigl(
R_{\uparrow\uparrow}\ket{\uparrow}
+R_{\downarrow\uparrow}\ket{\downarrow}
\Bigr)\otimes\ket{r_n}
\end{eqnarray}
and
\begin{eqnarray}
\ket{\downarrow}\otimes\ket{t_{n-1}}
&\to&\Bigl(
T_{\downarrow\downarrow}\ket{\downarrow}
+T_{\uparrow\downarrow}\ket{\uparrow}
\Bigr)\otimes\ket{t_n}\nonumber\\
&&{}+\Bigl(
R_{\downarrow\downarrow}\ket{\downarrow}
+R_{\uparrow\downarrow}\ket{\uparrow}
\Bigr)\otimes\ket{r_n},
\end{eqnarray}
respectively, where $T_{\downarrow\uparrow}$,
$T_{\uparrow\downarrow}$ ($R_{\downarrow\uparrow}$,
$R_{\uparrow\downarrow}$) are the probability amplitudes for
spin-flips when the neutron is transmitted (reflected), and the
two constraints
$|T_{\uparrow\uparrow}|^2+|T_{\downarrow\uparrow}|^2+
|R_{\uparrow\uparrow}|^2+|R_{\downarrow\uparrow}|^2=1$ and
$|T_{\downarrow\downarrow}|^2+|T_{\uparrow\downarrow}|^2+
|R_{\downarrow\downarrow}|^2+|R_{\uparrow\downarrow}|^2=1$ hold.
Hence the action of the magnetic mirror on a neutron in a general
spin state $\ket{S}$ reads
\begin{eqnarray}
\ket{S}\otimes\ket{t_{n-1}}
&\equiv&\Bigl(
c_\uparrow\ket{\uparrow}+c_\downarrow\ket{\downarrow}
\Bigr)\otimes\ket{t_{n-1}}\nonumber\\
&\to&\Bigl[
c_\uparrow\,\Bigl(
T_{\uparrow\uparrow}\ket{\uparrow}
+T_{\downarrow\uparrow}\ket{\downarrow}
\Bigr)\nonumber\\
&&\phantom{\Bigl[}
{}+c_\downarrow\,\Bigl(
T_{\downarrow\downarrow}\ket{\downarrow}
+T_{\uparrow\downarrow}\ket{\uparrow}
\Bigr)
\Bigr]\otimes\ket{t_n}\nonumber\\
&&{}+\Bigl[
c_\uparrow\,\Bigl(
R_{\uparrow\uparrow}\ket{\uparrow}
+R_{\downarrow\uparrow}\ket{\downarrow}
\Bigr)\nonumber\\
&&\phantom{{}+\Bigl[}
{}+c_\downarrow\,\Bigl(
R_{\downarrow\downarrow}\ket{\downarrow}
+R_{\uparrow\downarrow}\ket{\uparrow}
\Bigr)
\Bigr]\otimes\ket{r_n}\nonumber\\
&=&\tilde{\mathcal{T}}\ket{S}\otimes\ket{t_n}
+\tilde{\mathcal{R}}\ket{S}\otimes\ket{r_n},
\end{eqnarray}
where
\begin{subequations}
\begin{equation}
\tilde{\mathcal{T}}
=\ket{\uparrow}T_{\uparrow\uparrow}\bra{\uparrow}
+\ket{\uparrow}T_{\uparrow\downarrow}\bra{\downarrow}
+\ket{\downarrow}T_{\downarrow\uparrow}\bra{\uparrow}
+\ket{\downarrow}T_{\downarrow\downarrow}\bra{\downarrow},
\end{equation}
\begin{equation}
\tilde{\mathcal{R}}
=\ket{\uparrow}R_{\uparrow\uparrow}\bra{\uparrow}
+\ket{\uparrow}R_{\uparrow\downarrow}\bra{\downarrow}
+\ket{\downarrow}R_{\downarrow\uparrow}\bra{\uparrow}
+\ket{\downarrow}R_{\downarrow\downarrow}\bra{\downarrow}.
\end{equation}
\end{subequations}
Compare with Eq.\ (\ref{eqn:PseudoProjectionOp}). The operator
$\tilde{\mathcal{T}}e^{-iHt/N\hbar}$ reads now
\begin{eqnarray}
&&\tilde{\mathcal{T}}e^{-iHt/N\hbar}\nonumber\\
&&\ =\frac{1}{2}\left[
 (T_{\uparrow\uparrow}+T_{\downarrow\downarrow})
 \cos\frac{\theta}{N}
 -i(T_{\uparrow\downarrow}+T_{\downarrow\uparrow})
 \sin\frac{\theta}{N}
\right]\nonumber\\
&&\ \phantom{=}
{}+\sigma_x\frac{1}{2}\left[
 (T_{\uparrow\downarrow}+T_{\downarrow\uparrow})
 \cos\frac{\theta}{N}
 -i(T_{\uparrow\uparrow}+T_{\downarrow\downarrow})
 \sin\frac{\theta}{N}
\right]\nonumber\\
&&\ \phantom{=}
{}+\sigma_y\frac{i}{2}\left[
 (T_{\uparrow\downarrow}-T_{\downarrow\uparrow})
 \cos\frac{\theta}{N}
 -i(T_{\uparrow\uparrow}-T_{\downarrow\downarrow})
 \sin\frac{\theta}{N}
\right]\nonumber\\
&&\ \phantom{=}
{}+\sigma_z\frac{1}{2}\left[
 (T_{\uparrow\uparrow}-T_{\downarrow\downarrow})
 \cos\frac{\theta}{N}
 -i(T_{\uparrow\downarrow}-T_{\downarrow\uparrow})
 \sin\frac{\theta}{N}
\right]\nonumber\\
\end{eqnarray}
and its eigenvalues $\xi_\pm(N)$ are given by
\begin{subequations}
\label{eqn:EigenvaluesWithSpinFlip}
\begin{equation}
\xi_\pm(N)
=C\pm\sqrt{C^2-(T_{\uparrow\uparrow}T_{\downarrow\downarrow}
-T_{\uparrow\downarrow}T_{\downarrow\uparrow})} ,
\end{equation}
with
\begin{equation}
C=\frac{1}{2}\left[
(T_{\uparrow\uparrow}+T_{\downarrow\downarrow})
\cos\frac{\theta}{N}
-i(T_{\uparrow\downarrow}+T_{\downarrow\uparrow})
\sin\frac{\theta}{N}
\right].
\end{equation}
\end{subequations}
A calculation similar to that in Sec.\ \ref{sec:QZENonIdeal}
yields the survival probability
\begin{eqnarray}
&&\tilde{P}^{(N)}(\theta)\nonumber\\
&&\quad=\left| A(N)-B(N)\left(
 T_{\downarrow\downarrow}\cos\frac{\theta}{N}
 -iT_{\downarrow\uparrow}\sin\frac{\theta}{N}
\right)
\right|^2\nonumber\\
&&\quad\phantom{=}
{}+\left|
B(N)\left(
 T_{\downarrow\downarrow}\sin\frac{\theta}{N}
 +iT_{\downarrow\uparrow}\cos\frac{\theta}{N}
\right)
\right|^2,
\label{eqn:PNonIdealWithSpinFlip}
\end{eqnarray}
where $A(N)$ and $B(N)$ are defined as in Eqs.\ (\ref{eqn:A})
and (\ref{eqn:B}), respectively, but with the eigenvalues
$\xi_\pm(N)$ in Eqs.\ (\ref{eqn:EigenvaluesWithSpinFlip}). For
$|T_{\downarrow\downarrow}|$, $|T_{\uparrow\downarrow}|$,
$|T_{\downarrow\uparrow}|\ll|T_{\uparrow\uparrow}|$, the
probability (\ref{eqn:PNonIdealWithSpinFlip}) is readily evaluated
as
\begin{eqnarray}
\tilde{P}^{(N)}(\theta)
&\simeq&|T_{\uparrow\uparrow}|^{2N}
\left(\cos\frac{\theta}{N}\right)^{2N}\nonumber\\
&&{}\times\biggl[
1-2\Re\!\left(
\frac{T_{\downarrow\downarrow}}{T_{\uparrow\uparrow}}
\right)(N-1)\tan^2\!\frac{\theta}{N}\nonumber\\
&&\phantom{{}\times\biggl[1}
{}+2\Im\!\left(
\frac{T_{\uparrow\downarrow}}{T_{\uparrow\uparrow}}
\right)N\tan\frac{\theta}{N}\nonumber\\
&&\phantom{{}\times\biggl[1}
{}+2\Im\!\left(
\frac{T_{\downarrow\uparrow}}{T_{\uparrow\uparrow}}
\right)(N-1)\tan\frac{\theta}{N}
+\cdots
\biggr],\nonumber\\
\end{eqnarray}
which shows that the probability $\tilde{P}^{(N)}(\theta)$ is
again dominated by the factor (\ref{eqn:PApp}) (with
$|T_\uparrow|$ replaced by $|T_{\uparrow\uparrow}|$), and the
spin-flips at the mirrors yield only a first-order correction.

\end{document}